\documentclass{PoS}

\usepackage{natbib}

\title{Probing First Galaxies and Their Impact on the
  Intergalactic Medium through the
  21-cm Observation of the Cosmic Dawn  with the SKA}

\ShortTitle{First Galaxies}

\author{\speaker{Kyungjin Ahn}\\
        Department of Earth Sciences, Chosun University, Gwangju
        501-759, Korea\\
        E-mail: \email{kjahn@chosun.ac.kr}}

\author{Andrei Mesinger\\
        Scuola Normale Superiore, Piazza dei Cavalieri 7, 56126 Pisa, Italy\\
        E-mail: \email{andrei.mesinger@sns.it}}

\author{Marcelo A. Alvarez\\
        Canadian Institue for Theoretical Astrophysics, 60 St. George St., Toronto, ON. M5S 3H8, Canada\\
        E-mail: \email{malvarez@cita.utoronto.ca}}

\author{Xuelei Chen\\
        National Astronomical Observatory of China, 20A Datun Road, Chaoyang District, Beijing, China, 100012 \\
        E-mail: \email{xuelei@cosmology.bao.ac.cn}}

\abstract{We present an overview of the theory of high-redshift star and
  X-ray source 
  formation, and how they affect the 21-cm background. Primary focus is
  given to Ly$\alpha$ pumping and 
  X-ray heating mechanisms at cosmic dawn, opening a new observational window
  for high-redshift astrophysics by generating sizable fluctuations in
  the 21-cm background. We describe observational prospects for 
  power spectrum analysis and 3D tomography (imaging) of the signature of
  these early astrophysical sources by SKA1-LOW and SKA2.}

\FullConference{
Advancing Astrophysics with the Square Kilometre Array\\
June 8-13, 2014\\
Giardini Naxos, Italy}

\def\aap{{ A\&A}}

\def\apj{{ ApJ}}
\def\apjl{{ ApJL}}

\def\mnras{{ MNRAS}}
\def\nat{{ Nature }}

\def\prd{{ Phys.\ Rev.\ D }}

\def\dtb{\delta T_{b}}

\begin{document}

\section{Introduction}
\label{sec:intro}
The early phase of the Epoch of Reionization (EoR) or Cosmic Dawn
(CD) is, by definition, believed to have started with the formation
of the first stars, even though the EoR may also have been fuelled initially
by somewhat unconventional radiation sources such as annihilating dark matter 
clumps (e.g. \citealt{Spolyar2009}). The first stars are believed to
have formed within minihalos (halos with $T_{{\rm vir}}\lesssim10^{4}\,{\rm K}$,
or $10^{4}\lesssim M/M_{\odot}\lesssim10^{7-8}$) and to have been massive,
emitting a sufficient amount of UV radiation to ionize the surrounding hydrogen
and helium. These stars were born at zero metallicity, with their formation
predominantly regulated by ${\rm H_{2}}$ cooling,
and thus they are also Population III (Pop III) stars. Until recently the conventional view, from high-resolution numerical simulations, was of massive ($\gtrsim100\, M_{\odot}$)
stars forming in isolation (e.g. \citealt{2002Sci...295...93A}; \citealt{2002ApJ...564...23B};
\citealt{Yoshida2006}): one Pop III star per minihalo.
This paradigm has undergone recent revision due to newer simulations
in which smaller-mass ($\sim10-40\, M_{\odot}$) stars
form binary systems (\citealt{Turk2009}; \citealt{Stacy2010}; \citealt{Greif2011a}),
although the universality of this result is unclear (see \citealt{Hirano2014}
for a wide spectrum of Pop III stars in $\sim100$ minihalos). At
any rate, a change in the initial mass function (IMF) of Pop III stars
affects the hardness of the spectral energy distribution (SED) of these
stars, and thus the CD and EoR modelling as well. Stellar binaries
may evolve into X-ray binaries that may, if efficient, result in a smoother
ionization structure of the intergalactic medium (IGM) than that by UV
sources, due to a much longer mean 
free path than that of typical UV photons (e.g. 
\citealt{Haiman2011}; \citealt{Mesinger2013} and references therein). Furthermore, the
interstellar medium heated by supernova explosions can cool through a combination of
Bremsstrahlung and metal line cooling, producing soft ($\lesssim$keV) X-rays that can
efficiently heat the high-redshift IGM (\citealt{Pacucci2014}).

The theory of the formation and evolution of the first stars, therefore,
is crucial in modelling CD and even the EoR, because the metallicity
built up after the death of Pop III stars will give way to
the formation of Population II (Pop II) stars, believed
to be the main drivers of reionization. A major feedback that affects
their formation is the dissociation of ${\rm H}_{2}$ molecules --
cooling agents -- inside minihalos by Lyman-Werner band radiation
(e.g. \citealt{Haiman2000}; for recent simulations with this effect
in $\gtrsim100$ Mpc box see e.g. \citealt{Ahn2012} and \citealt{Fialkov2013}).
Stars would not have formed either in minihalos that were embedded in photoionized gas
($T\gtrsim10^{4}\,{\rm K}$),
since the increase in pressure would have suppressed minihalo gas accretion (Jeans-mass filtering).

There also appeared a noteworthy discovery having to do with the nature of large scale fluctuations. \citet{Tseliakhovich2010} used a peak-background split
scheme to study the nonlinear evolution of the baryon/dark-matter velocity
offset, which was seeded at the recombination epoch, and found that hydrodynamics
at scales relevant to minihalo formation should have been affected
(see more details of its astrophysical impact in the subchapter by Maio). The
typical offset is found to be e.g. $\sim1$ km/s at $z\sim20$,
corresponding to $M_{{\rm halo}}\sim1.5\times10^{5}\, M_{\odot}$,
when minihalos were already abundant.
While determination of the halo mass and environmental dependence of this effect awaits more detailed study, simulations already show an increase in the threshold mass for star formation
and a delay in halo collapse (\citealt{Stacy2011};
\citealt{Greif2011}; \citealt{OLeary2012}).
More importantly, this nonlinear effect may shock-heat the gas
globally, introducing a new 
feature into the power spectrum that corresponds to the 
relative velocity fluctuations (\citealt{McQuinn2012}). When the
global shock-heating is efficient, this new
feature could even dominate over the power spectrum of linear density
perturbations and amplify the baryon acoustic oscillation (BAO)
feature. Due to the uncertainties in the efficiency of this
shock-heating, however, the actual amplitude of the power spectrum due
to relative velocity fluctuations is still uncertain.

From an observational perspective on CD and the EoR, especially
in terms of 21-cm power spectrum analysis, it is convenient to
mark three prominent epochs: (1) the Ly$\alpha$-pumping epoch, when
the IGM is strongly coupled to $T_{{\rm K}}$ through Wouthysen-Field
effect with a high Ly$\alpha$ intensity, (2) the X-ray heating epoch,
when the IGM is gradually heated to beyond $T_{{\rm CMB}}$ by  X-ray heating
and (3) the EoR, when H II bubbles in cosmological scales form in a patchy
way. It is generally believed that CD commences with the  Ly$\alpha$-pumping
epoch, followed by the X-ray heating epoch, and finally occurs the
EoR, whose sequence is rather robust unless 1-2 order-of-magnitude
changes in the fiducial astrophysical
parameters are allowed (\citealt{2006MNRAS.371..867F};
\citealt{McQuinn2012}; \citealt{Mesinger2014}). Very conveniently, the
early phase of each epoch boosts the spatial fluctuation of
$\dtb$, sequentially dominated by the patchiness in the Ly$\alpha$
intensity, the IGM temperature, and the ionized fraction,
respectively, extending the observational window to very
high redshifts.

While challenging, it should be possible to observe individual objects via tomography,
which may be as small as the first galaxies. When UV sources are embedded
in an IGM colder than the CMB, a ``Ly$\alpha$ sphere'' forms around the
radiation source and a strong 21-cm absorption trough forms. When UV
sources are accompanied by X-ray sources, the central, heated region
is seen in emission, which is still surrounded by an absorption trough
or an absorption plateau (e.g. \citealt{2000ApJ...528..597T}; \citealt{Cen2006};
\citealt{2006ApJ...648L...1C}; \citealt{Chen2008}; \citealt{Alvarez2010}).
A simulation of the formation and evolution of Pop III and II stars
in a rare density peak, together with the X-ray binaries, shows that
this ubiquitous feature will be observable by SKA2-LOW with 1000-hour
integration, and marginally 
by SKA1-LOW (\citealt{Ahn2014}) with very aggressive integration at
$z\lesssim 15$ (Section~\ref{sec:imaging}).  

\section{Power Spectrum Analysis}
\label{sec:pk}

Because many independent modes with similar wavenumbers ($k\equiv|{\bf k}|$)
are averaged over to form the 3D
power spectrum, $P(k)$, power spectrum analysis
has higher sensitivity than imaging in general.
In addition,
nature cooperates in such a way that three important physical
processes -- Ly$\alpha$ pumping, X-ray heating and patchy
reionization -- may sequentially boost 21-cm fluctuations to give a roughly constant S/N ratio over redshift, even though the
foreground increases rapidly towards high redshift (e.g. \citealt{Mesinger2014}). Overall, 
the signal and the foreground noise grow roughly at a similar rate to yield S/N$\sim 10^{-5}$, although the S/N ratio should eventually become very small ($\sim 10^{-7}$) at $z\sim 30$ \citep{Pritchard2008}.

\begin{figure}
\vspace{-3cm}
\includegraphics[width=0.54\textwidth]{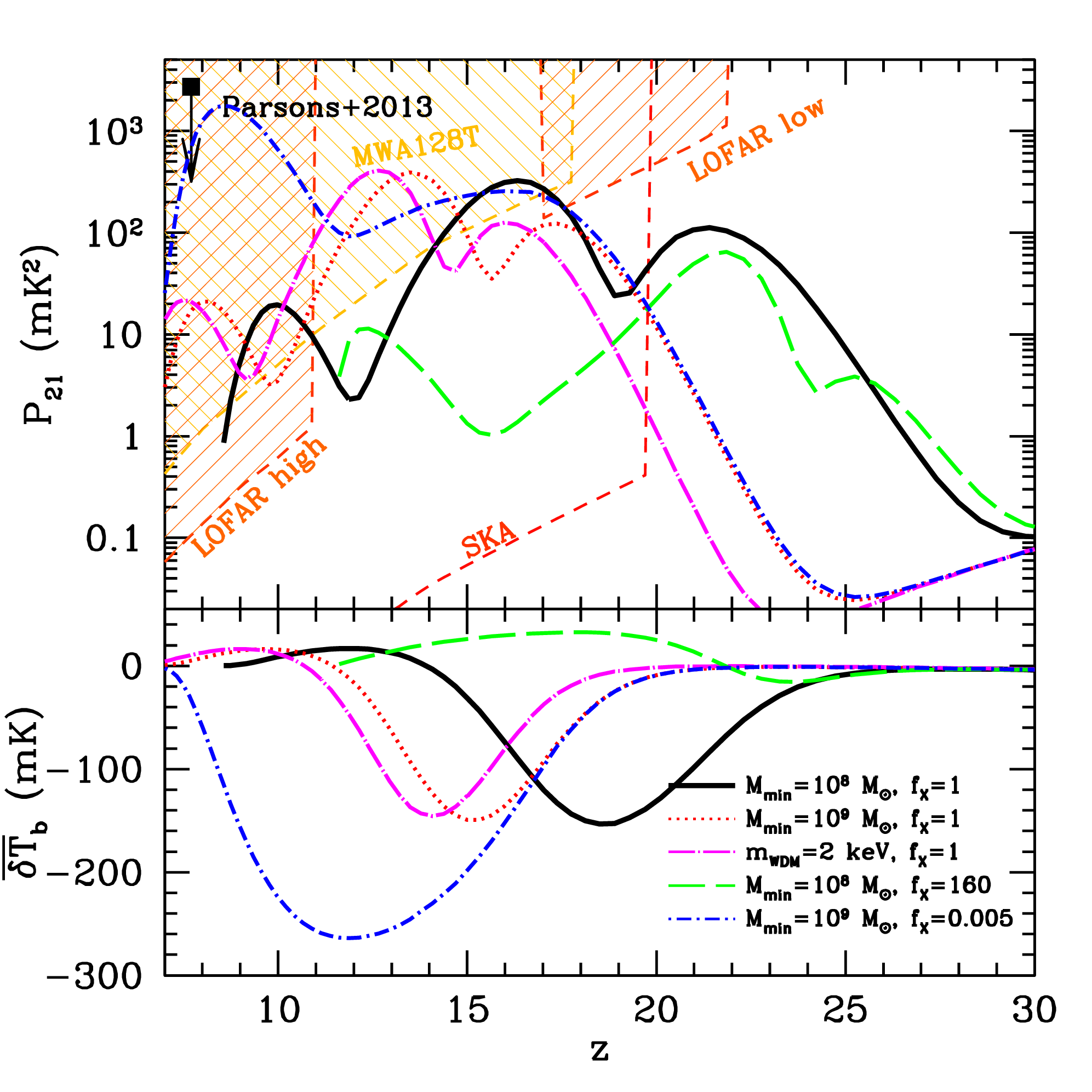}
\hspace{-0.8cm}
\includegraphics[width=0.54\textwidth]{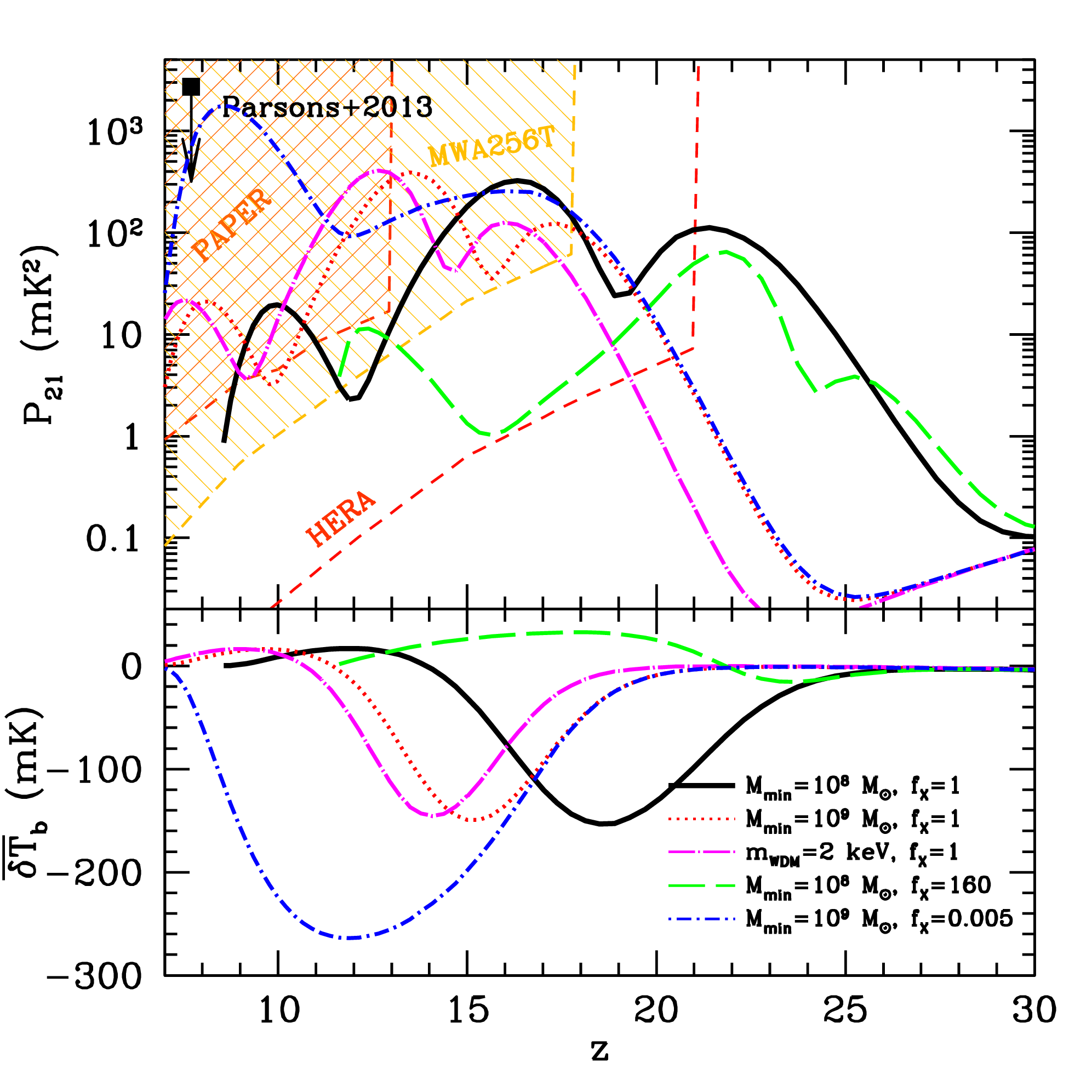}
\caption{(top subpanels) Evolution of $D^{2}(k)$ at $k=0.1/{\rm Mpc}$ and sensitivity
of various planned observations including SKA1-LOW (We note that the
apparent SKA1-LOW sensitivity cut at $z\approx20$ is 
due to the particular choice of instantaneous bandwidth, $\Delta
z=0.5$, which limits the number of available $k_{||}=0.1/{\rm Mpc}$
modes; we stress that bandwidth choices are arbitrary and that
SKA1-LOW's planned bandpass extends out to $z=28$.). 
Three peaks appear when fluctuations
in $\dtb$ are boosted due to the efficient Ly$\alpha$-pumping,
the X-ray heating and the patchy reionization, subsequently from high to low
redshifts (\citealt{Mesinger2014}).}

\label{fig:mesingerPk}

\end{figure}

\begin{figure}
\begin{center}
\includegraphics[width=0.5\textwidth]{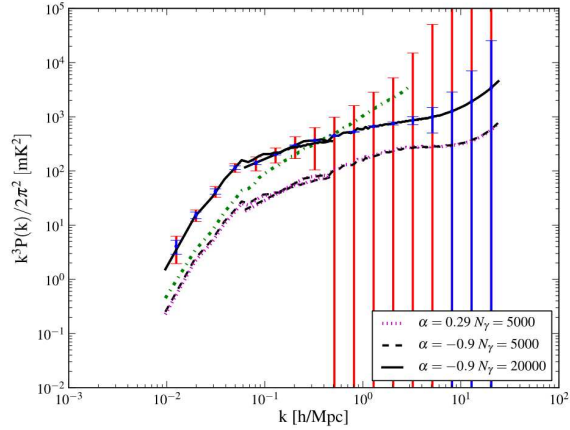}
\end{center}

\caption{21-cm temperature power spectrum  at $z\simeq 20.3$ for a
  few Ly$\alpha$ emission models, 
  based on the simulation results by \citet{Santos2011}. Power-law
  SEDs are used: the number of photons per frequency emitted per
  stellar baryon $\epsilon_{\nu}=A\nu^{\alpha}$, where $A$ is tuned to
  produce the assumed integrated
  number of photons between Ly$\alpha$ (10.2 eV) and the Lyman limit
  (13.6 eV), or $N_{\gamma}$ (legends are self-explanatory).
  $\alpha=-0.9$ and $\alpha=0.29$ roughly represent Pop-II and Pop-III
  type SEDs, respectively. Cases with identical $N_\gamma$ (black solid and
  black dashed curves) are almost degenerate.
  Pushing the minimum halo mass hosting stars down 
  to $M_{\rm min}=10^{6}M_\odot$, with $\alpha=0.25$ and $N_\gamma=5000$,
  renders the overall shape of the power spectrum different from the
  other three cases, which all have $M_{\rm min}=10^{8}M_\odot$.  
Error bars in blue corresponds to the full SKA2-LOW collecting area, and the
ones in red to the 10\% of the collecting area
(\citealt{Santos2011}).}

\label{fig:santosPk}

\end{figure}

The very first light from astrophysical sources
to which the IGM is exposed is the stellar continuum below the hydrogen Lyman limit, or
photons with energy $h\nu<13.6\,{\rm eV}$, because ionizing
photons (photons above the Lyman limit) are quickly absorbed by
the neutral hydrogen and helium around while continuum photons are not. However, there
is always some fraction of this continuum that is absorbed in Lyman series
resonances as photons redshift over cosmological distances (e.g. \citealt{Ahn2009}). After
absorption, cascades over the energy levels and 
multiple scattering of reemitted photons will convert about $\sim 30\%$ 
of those absorbed photons to Lyman-$\alpha$ photons
(\citealt{Hirata2006,Pritchard2006}). Even in the presence of density inhomogeneities, a point
source produces a nearly spherical profile
(\citealt{Vonlanthen2011}) that shows a step-wise feature decreasing more steeply than $1/r^2$ (\citealt{Pritchard2006}). As usually
assumed, the IGM is still much colder than the CMB, 
and thus the regions where the Ly$\alpha$ pumping becomes strong
($N_{\alpha}\gtrsim N_{\alpha, {\rm th}} \simeq
10^{-10}\left(\frac{20}{1+z}\right)$~${\rm s}^{-1} {\rm cm}^{-2} {\rm
  Hz}^{-1} {\rm sr}^{-1}$) will show strong absorption, or
$\dtb \sim -100\,{\rm mK}$. Source clustering and the steep radial
profile of the
Ly$\alpha$ intensity around each source are the dominant sources of
fluctuations in $\dtb$ at earliest times. Soon thereafter, saturation occurs, and fluctuations decrease as sources grow in abundance and $N_{\alpha}\gg 1$ everywhere
(Fig.~\ref{fig:mesingerPk}). Changes in the
spectral energy distribution (SED) of sources can affect the resulting
power spectrum as demonstrated by \citet{Santos2011}
(Fig.~\ref{fig:santosPk}), while other 
systematic effects such as the star formation efficiency might make
the prediction somewhat degenerate. 

In many models, the Ly$\alpha$ pumping epoch is followed by the
X-ray heating epoch. X-ray photons have a much longer mean free path
than UV photons, and partially ionize the IGM by leaving behind energetic
electrons as they traverse cosmological distances. These electrons
can then ionize, excite, and heat the IGM. During the early phase of the
X-ray heating epoch, the overall ionization level caused by X-rays is usually very
small unless the X-ray emissivity is very high (see e.g. \citealt{Mesinger2013}). Thus the
dominant contribution from X-rays to fluctuations in $\dtb$ is
inhomogeneous heating, which will boost the power spectrum again (Fig.~\ref{fig:mesingerPk}). Another contribution to fluctuations is due to inhomogeneous Ly$\alpha$ pumping from excited 
hydrogen atoms (see e.g. \citealt{Chen2008}), though its
contribution is usually subdominant compared to the Ly$\alpha$ pumping induced
by stellar UV photons (but see e.g. \citealt{Ahn2014} for a significant
contribution when X-ray source formation is very efficient). It is
usually expected that X-ray heating epoch produces the highest
peak in the large-scale power spectrum in its redshift evolution
(Fig.~\ref{fig:mesingerPk}). Later, 
X-ray heating will become very efficient everywhere in the Universe
such that $\dtb>0$ and $T_{\rm IGM}\gg T_{\rm CMB}$ but with the
ionization state still low, once again saturating the fluctuations and resulting in a significant decrease in the amplitude of the power spectrum (Fig.~\ref{fig:mesingerPk}). This rise, peak and fall in the
power spectrum amplitude results in the second ``bump'' seen in figure~\ref{fig:mesingerPk}. Note that
this provides the best window for high-z cosmology, because $\dtb$
becomes proportional to the baryon density and nearly insensitive to other
fluctuations. In addition, separating the matter power spectrum from
that of the radiation fields, utilizing the impact by the peculiar motion
(or the so-called $\mu$-decomposition scheme which is very similar to that
used in galaxy survey programs),
may be possible (\citealt{2006ApJ...653..815M}; \citealt{Mao2012};
see also the subchapter by Pritchard).

Finally comes the main EoR, in most models. The power spectrum is boosted again, due
to the patchiness of H II regions where $\dtb=0$ and neutral regions
where $\dtb$ is proportional to the baryon density. Final saturation and power spectrum suppression occurs as bubbles overlap and the EoR approaches its final phase (e.g. \citealt{selfregulated:new}), to generate
the final bump in the power spectrum evolution (Fig.~\ref{fig:mesingerPk}). Of 
course, one should allow more model variants. Currently, the X-ray
emissivity at CD and during the EoR is still largely unknown, other than the
weak constraint coming from the soft X-ray background
(e.g. \citealt{2004ApJ...613..646D}; \citealt{McQuinn2012a}), the $z\sim
2.5$ QSO absorption lines and the reionization histories of H I
and He II (\citealt{McQuinn2012a}). This allows a wide model variation in the
X-ray heating epoch, and thus the 3-bump feature in the power spectrum
(Fig.~\ref{fig:mesingerPk}) is not ubiquitous among reionization
models if extreme cases are allowed (\citealt{Pritchard2012}).
The halo mass spectrum responsible for reionization is not well
constrained either, allowing models with a significant contribution from
minihalos (e.g. \citealt{Wyithe2007}; \citealt{Ahn2012}).

The sweet spot for power spectrum analysis exists around $k\sim
0.1/{\rm Mpc}$, guaranteeing a large S/N over a wide range of
redshifts while being mostly unaffected by the filtering of low-$k$ foreground modes. SKA1-LOW (and the 50\% capability of SKA1-LOW as well) has
a large enough S/N ratio for observing $P(k\sim 0.1/{\rm Mpc})$ for
$z\lesssim 20$ with 1000-hour integration, and SKA2-LOW will allow one
to achieve similar S/N 
ratio for even larger $k$'s with more extended redshift coverage,
$z\lesssim 28$
(Fig.~\ref{fig:mesingerPk}; see also 
\citealt{Baek2010} for the evolution of other wavemodes). As is
usually the case for CD and the EoR, pushing the observable redshift to
$z\sim28$ or $\nu\sim 50 {\rm MHz}$ is strongly favored, in order to probe their
early phases.

\section{Tomography}
\label{sec:imaging}

As mentioned in Section~\ref{sec:pk}, imaging is more difficult than
power spectrum analysis. While the power spectrum contains a lot of
information on the underlying astrophysics and cosmology, one should
note that the 3D tomography (imaging) is indeed a more direct
observation because one cannot invert $P(k)$ to obtain the actual image
unless the 21cm field is Gaussian (see also the subchapter
by Mellema).

\begin{figure}
\includegraphics[width=0.48\textwidth]{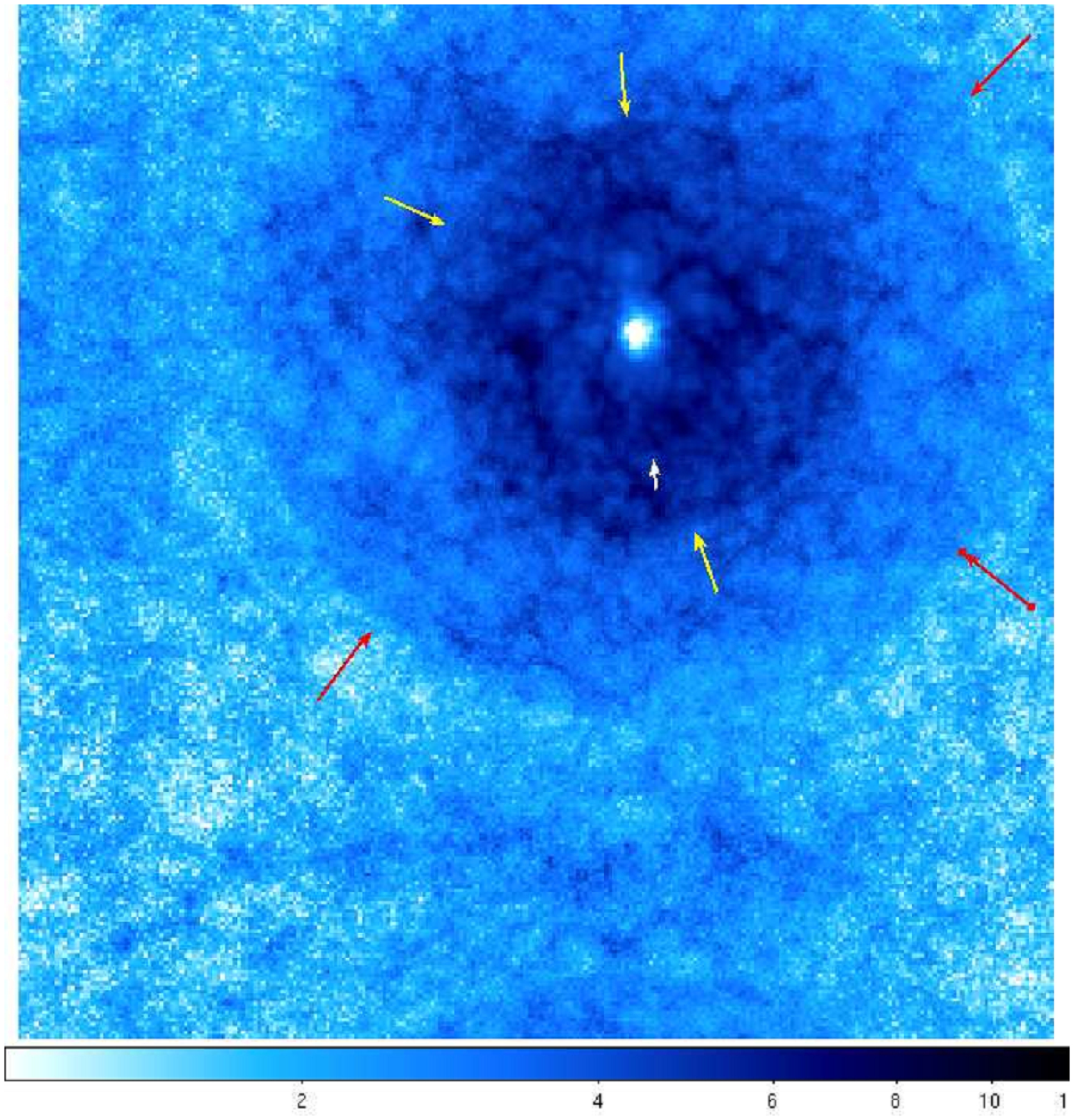}
\includegraphics[width=0.52\textwidth]{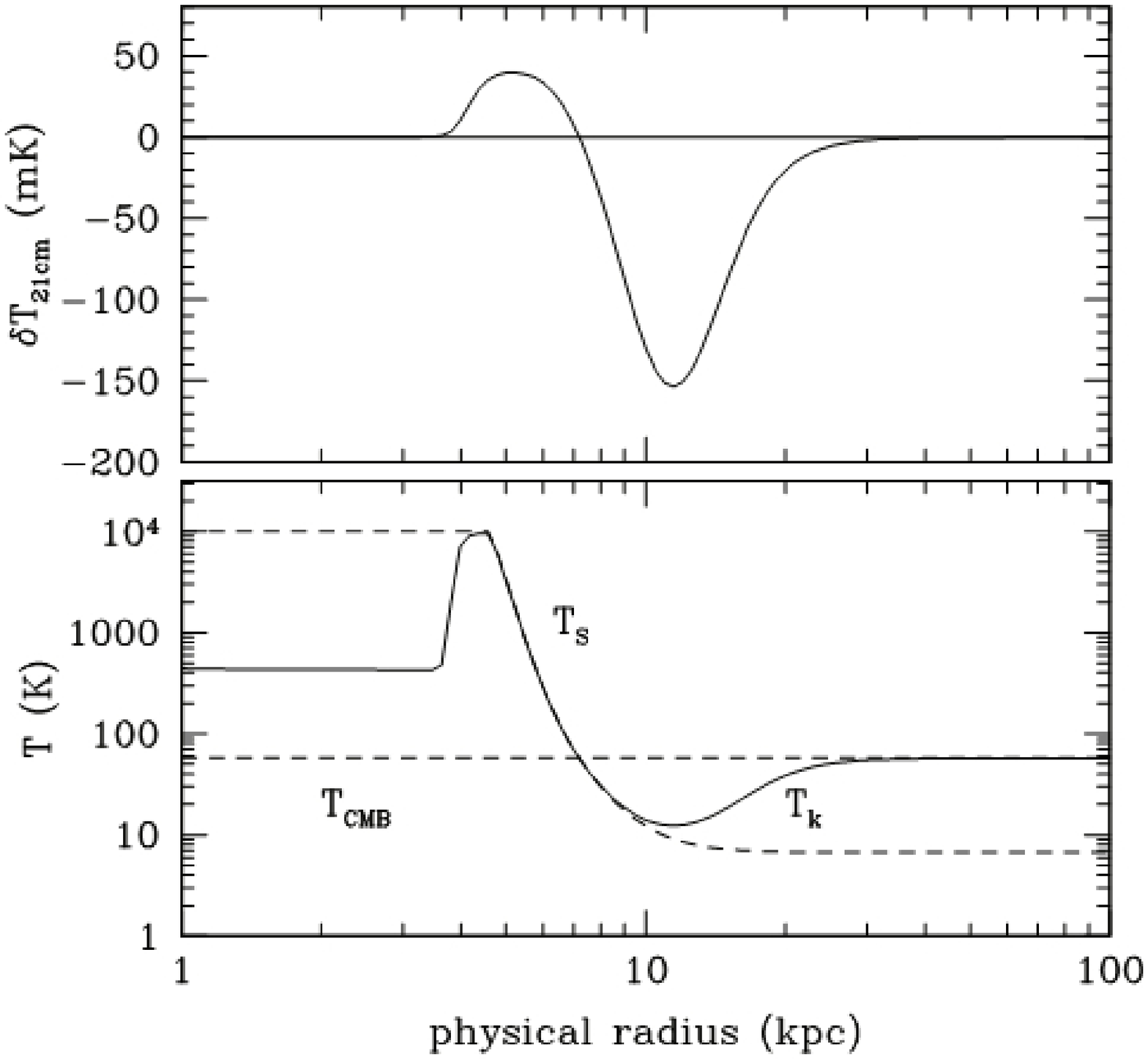}

\caption{(a:left) Distinct ring structure due to the step-wise
  Ly$\alpha$ pumping around 
a point source. $-\dtb \times r^{2}$ at $z=13.42$ is plotted with an
arbitrary unit, to show these distinct rings (pointed
by arrows) as boundaries of sudden change in color: each annulus bound
by adjacent rings, where Ly$\alpha$ flux decreases as $1/r^2$, shows
about the same color. The box size is 137 Mpc coming and the angular
size is $51'$ (\citealt{Vonlanthen2011}).
(b:right) Radial profile of an isolated radiation source of the UV and the
X-ray. From inside out, $\dtb=0$ (H II region), $\dtb>0$ (X-ray heated
region), $\dtb<0$ (Ly$\alpha$-pumped region).}

\label{fig:Lyring}

\end{figure}

\begin{figure}

\begin{center}
\includegraphics[width=0.4\textwidth]{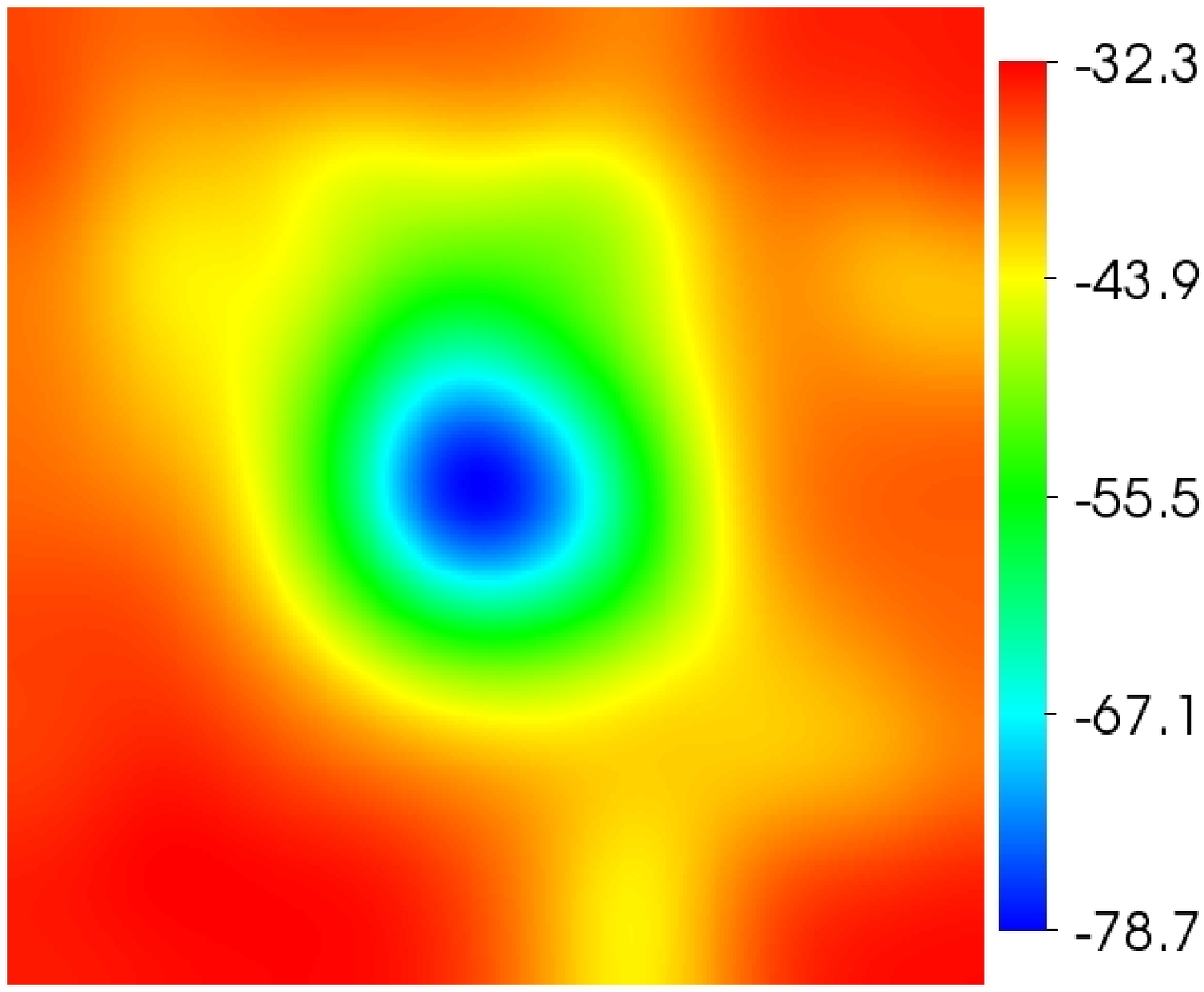}
\includegraphics[width=0.4\textwidth]{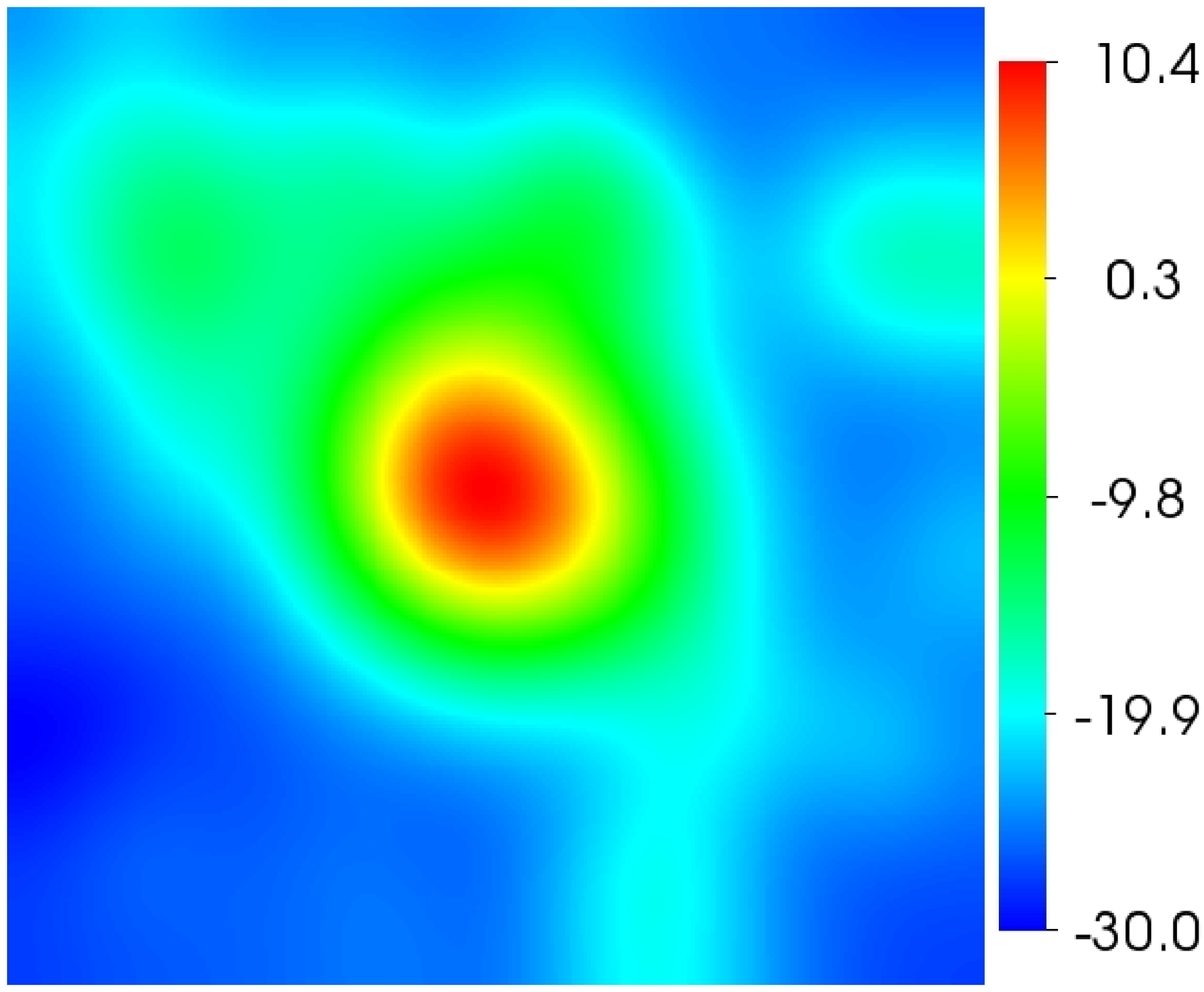}
\vskip 1em
\includegraphics[width=0.6\textwidth]{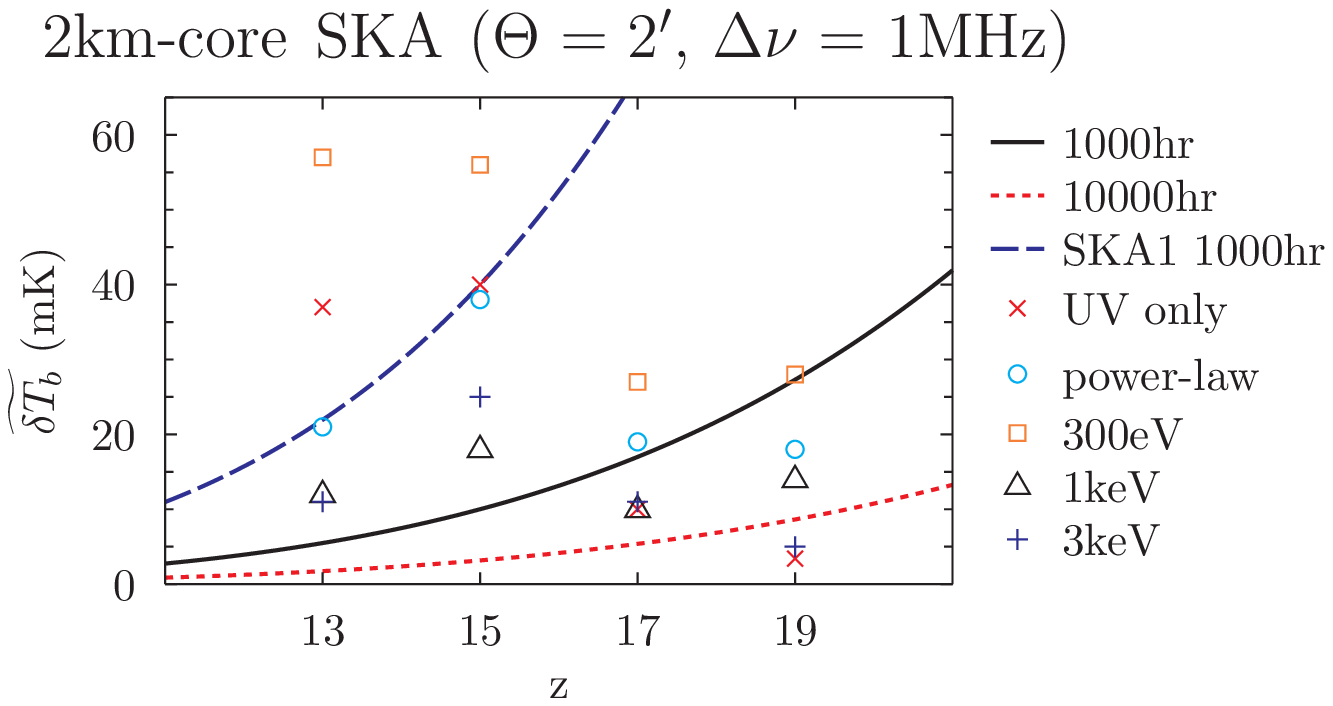}
\end{center}
\caption{Realistic imaging forecast of the early objects with the
  first stars (Population III stars), Population II stars and
  Population III X-ray binaries, (\citealt{Ahn2014}).
  (a:top left) 21cm map (mK) from a simulated model of Pop
  III stars+Pop II stars {\em without} X-ray sources in a rare density
  peak  inside a 40 Mpc (comoving) box at $z=15$, smoothed with
  $\Theta=2'$ and $\Delta\nu=1\,$MHz. The angular size of the box
    shown is $13'$ at this redshift.
  (b:top right) 21cm map (mK) from a simulated model of Pop
  III stars+X-ray binaries+Pop II stars in a rare density
  peak  inside the same box as (a) at $z=15$, smoothed with
  $\Theta=2'$ and $\Delta\nu=1\,$MHz. 
  (c:bottom) Imaging  sensitivity of the 2-km core
  SKA2-LOW (\citealt{Mellema2013}) with $\Theta=2'$ and $\Delta\nu=1\,$MHz,
  for integration times of 1000 hours (black, solid curve) and 10000 hours
  (red, dotted curve), against varying X-ray binary
  model predictions (points). It is difficult to image individual
  objects by SKA1-LOW unless nature cooperates: a few models
  are marginally observable at $z\lesssim 15$, as seen by 
  the sensitivity of  SKA1-LOW with the same smoothing filters and
  a 1000-hour integration time (blue, dashed curve).}

\label{fig:imaging}

\end{figure}

Imaging can probe physical properties of radiation sources. One
interesting feature is that due to the step-wise decrease of
Ly$\alpha$ intensity (or Ly$\alpha$ pumping) in addition
to the geometrical dilution, $1/r^2$, the radial profile of the 21-cm
signal around an 
isolated source will also exhibit a step-wise variation, or multiple ``rings''
(\citealt{Vonlanthen2011}; Fig.~\ref{fig:Lyring}a). Detecting the
multiple-ring feature from one radiation
source remains, however, very difficult even with SKA2-LOW, and while
stacking many such profiles might allow the detection of the rings, one
should know the source redshift accurately, because stacking sources from
other redshift planes will quickly smear out the ring feature.

A more notable and traceable feature from an isolated source is (unless
the IGM is saturated $T_{\rm S}\gg T_{\rm CMB}$ due to strong
X-ray heating) $\dtb$ varying even more steeply than $1/r^2$. This is due to the step-wise variation of Ly$\alpha$ pumping
rate, resulting in strong absorption nearby and weak absorption further away. This
``Ly$\alpha$ blob'' should be ubiquitous around isolated sources
before the full saturation of Ly$\alpha$ pumping occurs everywhere,
which is analogous to individual H~II regions in the patchy reionization process
before the full overlap occurs (Fig.~\ref{fig:Lyring}a).

It becomes even more interesting when the UV source is accompanied by
an X-ray source. Typical emissivity of X-ray sources associated with
UV (stellar) sources is relatively small (compared to proposed
Population III X-ray binaries), and the corresponding X-ray heating
zone (where $T_{\rm S}>T_{\rm CMB}$) is smaller than the Ly$\alpha$
blob when the IGM were colder than the CMB before exposure. In addition, the
central region 
will be ionized mostly by UV sources, such that $\dtb$ is zero at the
center, positive outside, and negative even further outside
(Fig.~\ref{fig:Lyring}b). Note that probing the signature of individual objects
is likely to be possible in only a relatively narrow range of frequency
(redshift), since rapid growth in the abundance of galaxies will soon wash out
such a signature.

This can occur even when there is a strongly
clustered set of sources: \citet{Xu2014} simulated the formation of
Population III and II stars and X-ray binaries inside a high-density
peak (``Rarepeak'') in the early Universe, and \citet{Ahn2014}
postprocessed the simulated data and obtained a 21-cm map which shows
a much more extended $\dtb$ profile than that of a single
object (Fig.~\ref{fig:imaging}a, b). A similar prediction for the spatially
extended profile also exists 
for high-z QSO systems, very sparsely spaced in the Universe
(\citealt{2000ApJ...528..597T}). In case of the Rarepeak, SKA2-LOW performing
1000-hour integration of a tracked field with the $\Theta=2'$ and $\Delta\nu=1
{\rm MHz}$ smoothing filter will
enable its (their) detection regardless of the model variation, ranging from
the all-absorption trough case (without X-ray binaries) to the fully
X-ray heated case (with efficient X-ray heating) with a high S/N ratio
(Fig.~\ref{fig:imaging}c).

As
it is easy to estimate the number density of such peaks a priori, one can
forecast the odds to find those objects inside a dedicated field of
observation. For Rarepeak, which is a $3.5\,\sigma$ density peak when
the linear density field is filtered at $R\sim 3\,{\rm Mpc}$ comoving
scale, its number density is roughly $10^{-6}\,{\rm Mpc}^{-3}$. 
A tracking volume of FOV(SKA1-LOW)$\times (110-75)\,{\rm MHz} \simeq
5^{\circ}\times 5^{\circ} \times 35 \,{\rm MHz}$ (where FOV denotes the
field of view), which covers the
redshift range $z=[12-17]$, will host about 600 of those high-density
peaks. Thus detecting the Ly$\alpha$-blob feature of these objects at
$z\lesssim 15$ through the SKA2-LOW 1000-hour integration is bound to
succeed, while with the SKA1-LOW array configuration, about a
16000-hour integration is required to achieve the same
sensitivity. Alternatively, we may target the few cases which can be marginally
detected by SKA1-LOW with a 1000-hour integration with the same
smoothing scheme (Fig.~\ref{fig:imaging}).
While lower-redshift detections become much easier, at that time the distinct isolated feature may be erased by signals from other, more abundant density peaks if similar emissivity is assumed. In either case, since the exact epoch when this occurs is model-dependent, it is necessary to explore a wider range of the astrophysical parameter space than has been
investigated by \citet{Ahn2014} and still consider the possibility to observe
individual objects at lower redshifts by SKA1-LOW. It is also
  advisable to carry out imaging with SKA1-LOW even when it has
  reached 50\% capability of its final phase, because again there may
  exist objects whose size and signal strength are large
  enough to be successfully imaged. For example, AGN+galaxy systems
  \citep{2000ApJ...528..597T} or even rarer peaks than have been
  studied by \citet{Ahn2014} may appear in such an imaging
  observation, although a relatively low-redshift range, $z\simeq
  13$, may be adequate due to the low sensitivity expected.

Aside from the high-redshift nature of these early stars/galaxies, the
unique feature of the composite H II region + X-ray heated region +
Ly$\alpha$ blob is not something one expects from the main EoR
imaging. The main EoR imaging will focus mostly on large H II bubbles 
($R \gtrsim 20 \,{\rm Mpc}$) with $\dtb=0$ before filtering, and the
neutral region will simply fluctuate according to the underlying
density fluctuation. If so, tomography will tell us about source
properties during the peak of EoR (see the subchapter by Mellema and
the subchapter by Iliev), and cosmological information during
the early phase of EoR (see the subchapter by Pritchard).

\section{Conclusion}
\label{sec:conclusion}
We have briefly reviewed theories of high-redshift astrophysics and
their observational aspects during CD, when the first galaxies formed.
The first galaxies significantly affect the IGM, leading to signatures that will be
detectable through the observation of the 21-cm background. With new developments in the theory of first-star formation,
the number of feasible models of this early epoch is ever increasing.
These new developments include the initial mass function of Pop III
stars, the spectral energy distribution of stars and X-ray sources,
the byproduct of stars, and dynamical/radiative feedback effects, to
name a few. SKA1-LOW and SKA2-LOW have the capability of probing first
galaxies and their impact on the IGM through power spectrum
analysis and imaging.

SK1-LOW will be able to observe the signature of these objects mostly
through the 21-cm power spectrum, with better sensitivity and
thus deeper target-redshift than most SKA precursors. Three prominent
epochs, which are the Ly$\alpha$ pumping epoch, X-ray heating epoch,
and the main EoR will 
be observed as three distinct evolutionary phases in the amplitude
of the 21-cm power spectrum in the wavenumber range around $k\sim 0.1/{\rm
  Mpc}$. The shape of the power spectrum in
a wide dynamic range of $k$ might be able to tell us about the
properties of the early objects, such as their SED.

SKA2-LOW will be able to carry out imaging, as well as the power
spectrum analysis at higher sensitivity than SKA1-LOW. Imaging will help
to constrain source properties much better than  power spectrum
analysis, because of model degeneracy and loss of information
are inherent in the power spectra. Clustered sources of UV and X-ray 
sources may 
show their unique signature of the composite absorption and emission
of 21-cm lines against the CMB, as long as their angular scale becomes large
enough to achieve high S/N with SKA2-LOW. For serendipity, however, it
is advised to carry out imaging also with SKA1-LOW.

There exists tension between high-redshift astrophysics and cosmology.
Even at CD, the 21-cm signal of an
astrophysical origin may swamp that of a cosmological
origin. Nevertheless, there exists the possibility that the cosmological
signal may be boosted significantly, if mechanical energy conversion
occurs on cosmological scales \citep{McQuinn2012}. In either case, the
best window for cosmology lies at the highest redshift range planned
for SKA ($z\sim 28$).



\end{document}